\documentclass[journal]{IEEEtran}
\usepackage{amsmath,amsfonts}
\usepackage{algorithmic}
\usepackage{amsmath}
\usepackage{array}
\usepackage{pdfpages}
\usepackage[caption=false,font=normalsize,labelfont=sf,textfont=sf]{subfig}
\usepackage{textcomp}
\usepackage{stfloats}
\usepackage{url}
\usepackage{verbatim}
\usepackage{graphicx}
\usepackage{cite}
\usepackage{hyperref}
\usepackage{pgf}
\usepackage{makecell}
\usepackage{bbm}
\usepackage{multirow}

\newcommand{\BF}{}

\usepackage{xcolor}
\usepackage[color]{changebar}
\def\BibTeX{{\rm B\kern-.05em{\sc i\kern-.025em b}\kern-.08em
		T\kern-.1667em\lower.7ex\hbox{E}\kern-.125emX}}
\usepackage{balance}
\begin{document}

\title{Deep Learning based Positioning with Multi-task Learning and Uncertainty-based Fusion}

\author{\IEEEauthorblockN{Anastasios Foliadis\IEEEauthorrefmark{1}\IEEEauthorrefmark{2}, Mario H. Casta\~{n}eda Garcia\IEEEauthorrefmark{1}, Richard A. Stirling-Gallacher\IEEEauthorrefmark{1},  Reiner S. Thom\"a\IEEEauthorrefmark{2}}
	
	\IEEEauthorblockA{\IEEEauthorrefmark{1}\textit{Munich Research Center}, \textit{Huawei Technologies Duesseldorf GmbH}, 
		Munich, Germany \\
		\textit{\IEEEauthorrefmark{2}Electronic Measurements and Signal Processing}, \textit{Technische Universit\"at Ilmenau}, Ilmenau, Germany\\
		\{\href{mailto:anastasios.foliadis@huawei.com}{anastasios.foliadis}, 
		\href{mailto:mario.castaneda@huawei.com}{mario.castaneda}, 
		\href{mailto:richard.sg@huawei.com}{richard.sg}\}@huawei.com, 
		\href{mailto:reiner.thomae@tu-ilmenau.de}{reiner.thomae@tu-ilmenau.de}}}




\maketitle

\begin{abstract}

Deep learning (DL) methods have been shown to improve the performance of several use cases for the fifth-generation (5G) New radio (NR) air interface. In this paper we investigate user equipment (UE) positioning using the channel state information (CSI) fingerprints between a UE and multiple base stations (BSs). In such a setup, we consider two different fusion techniques: early and late fusion. With early fusion, a single DL model can be trained for UE positioning by combining the CSI fingerprints of the multiple BSs as input. With late fusion, a separate DL model is trained at each BS using the CSI specific to that BS and the outputs of these individual models are then combined to determine the UE's position. In this work we compare these different fusion techniques and show that fusing the outputs of separate models achieves higher positioning accuracy, especially in a dynamic scenario. We also show that the combination of multiple outputs further benefits from considering the uncertainty of the output of the DL model at each BS. For a more efficient training of the DL model across BSs, we additionally propose a multi-task learning (MTL) scheme by sharing some parameters across the models while jointly training all models. This method, not only improves the accuracy of the individual models, but also of the final combined estimate. Lastly, we evaluate the reliability of the uncertainty estimation to determine which of the fusion methods provides the highest quality of uncertainty estimates. 

\end{abstract}


\section{Introduction}
\label{sec:introduction}

\IEEEPARstart{A}{ccurate} user positioning is an enablers of several future services and technologies \cite{Zhou19,DeLima21,Wang22,Chen22} such as location-aware communication, vehicle to everything (V2X) applications, industrial internet of things (IIOT), cooperating robots, commercial applications, etc. For this purpose, radio-based positioning of user equipment (UE) in wireless communication networks can be considered \cite{Zafari19}. Multiple base stations (BSs) deployed in such networks allow the collection of channel state information (CSI) over distributed links, which can be exploited for positioning of a UE. 
The CSI consists of the channel across the spatial and frequency domain, where the large number of antennas and large available bandwidth of current and future communication networks \cite{Chen22}, e.g., fifth generation (5G) or upcoming sixth generation (6G), can provide a high angular and temporal resolution to enable high accuracy positioning.

Conventional radio-based positioning methods are generally model-based and usually follow a two-step approach. With CSI estimated at one BS \cite{Kakkavas19} or at multiple BSs \cite{Revisnyei23}, relevant parameters or measurements e.g., path delay, angle of arrival (AoA), reference signal receive power (RSRP), time difference of arrival (TDoA), etc, are first determined to subsequently compute the UE’s position in a second step. Recently, machine learning (ML) and artificial intelligence (AI)-based techniques have also been proposed for radio-based UE positioning\cite{Chen17, Yin17, Thomae18, Widmaier19, Butt21} which are primarily data-driven and not model-based. In particular, deep learning (DL) methods, particularly convolutional neural networks (CNNs) have shown promising results \cite{Wang17, Niitsoo18, Hsieh19, Foliadis21}, being able to achieve sub-meter accuracy. In such data-driven models, the CSI over subcarriers and antennas of a UE at a given position is considered as a fingerprint associated with the UE’s position. By leveraging the ability of wireless networks to collect large amounts of data, a database of CSI fingerprints associated with different UE’s positions along with the respective UE’s position label can be constructed. With the DL-based positioning methods, a neural network (NN) can be trained on a given database, such that afterwards the NN can be employed for estimating a UE’s position by providing the CSI of the UE as its input. Different types of fingerprints have been considered in the literature, including the received signal strength (RSS), the magnitude and/or phase of the CSI over subcarriers in the frequency domain and across antennas in the spatial domain \cite{Hsieh19, Foliadis21, Chapre14, Vieira17, Bast20}.

With the CSI of a UE available across multiple BSs, early fusion or late fusion can be considered for the DL-based positioning methods \cite{Niitsoo19,Foliadis2021ReliableDL}. In early fusion, the CSI fingerprints from multiple BSs are collected and bundled together to constitute a single CSI fingerprint associated with the UE’s position. Thus with early fusion, only one NN needs to be trained with a database comprising with fingerprints of the CSI across multiple BSs \cite{Niitsoo19}. On the other hand, with late fusion, one NN is assumed at each BS where the CSI is considered as a fingerprint of the UE’s location associated only with the given BS \cite{Foliadis2021ReliableDL}. The NN associated with that BS is trained with a database of CSI fingerprints from that BS, enabling the NN to determine the UE’s position based only on the CSI estimated by that BS. Afterwards, a final UE’s position estimate is obtained by combining the position estimates obtained by the NNs across the multiple BSs \cite{Foliadis2021ReliableDL, Goenueltas21}, e.g., with a weighted average.

The choice between early or late fusion generally depends on the application \cite{Ramachandram2017DeepML}. However, when considering changes in the UE-BS channel between the training phase and deployment phase, e.g., due to a blockage of the line of sight (LOS) between a UE and a BS, late fusion can benefit from uncertainty estimation \cite{Foliadis2021ReliableDL}. In particular, the NN at each BS can be trained to estimate the uncertainty in the UE’s position determined by each NN. This enables the late fusion approach, to determine the final position estimate for the UE considering the uncertainty of the multiple position estimates obtained across multiple BSs. In practice the most reliable position estimates have a larger impact in determining the final UE’s position. Uncertainty estimation can be computed based on simple approaches like Monte Carlo Dropout (MCD) \cite{Gal2016DropoutAA} and Deep Ensembles (DEs) \cite{Lakshminarayanan2017SimpleAS}, which characterize the uncertainty based on the variance of the positioning error obtained with multiple NNs, i.e., similar position estimates across the different NNs indicates lower uncertainty estimation. Uncertainty estimation methods have also been proposed in \cite{Stahlke2024UncertaintyBasedFM} and \cite{Salihu2021TowardsSU} to detect corrupted fingerprints. 

Most conventional approaches to positioning require a strong line-of-sight (LOS) path and may be
impaired in non-LOS (NLOS) conditions or when there is a strong multipath. Recent works such as \cite{Shahmansoori2017PositionAO, Kakkavas19} have shown how to take advantage of the multipath information for single anchor UE positioning but are limited to multiple-input multiple-output (MIMO) systems and require prior knowledge of the nature of the incoming paths (i.e., LOS or NLOS). On the other hand DL-based methods can still be employed in strong multipath scenarios and don’t require multiple antennas at both receiver and transmitter. Despite this fact, with the multipath profile being susceptible to environmental changes, a DL model trained with CSI fingerprints from one environment may achieve a poor performance for the UE positioning in another environment \cite{Stahlke22, Foliadis2021ReliableDL}. 

The lack of direct transferability of the knowledge acquired in one environment to other environments is one of the challenges of DL-based positioning \cite{Owfi23}. The most straightforward way to address this is to retrain the NN from scratch with CSI fingerprints from the new environment, which may however be resource expensive and may not always be feasible.The resource intensive nature of position labeling that is required can be reduced by employing channel charting \cite{Studer2018ChannelCL} and by considering distance metrics between CSI fingerprints to create a map of the deployment scenario\cite{Stahlke2022IndoorLW, Stephan2023AngleDelayPA} using no or very few position labels. On the other hand, several approaches can be considered for improving the generalizability of a trained model to adapt it to environmental changes or to a new environment including transfer learning, domain generalization, multi-task learning and meta learning. With transfer learning, a previously trained model is used as an initial model that is fine-tuned with reduced training data from a new environment \cite{Bast20,Stahlke22}, which allows to speed up the training and to improve the performance compared to training from scratch. 

Furthermore, with multi-task learning (MTL) the aim is to jointly learn multiple models by training them while also sharing some or all of their parameters, thereby benefiting from regularization \cite{Hospedales2021MetaLearningIN}. Consequently, by considering positioning in different environments as different tasks, the positioning across multiple environments can be improved. When training a MTL scheme the choice of the relative importance of each task has to be considered. The hardest to learn tasks should be weighted less, so that the model focuses more on tasks that are easier to learn. Based on the uncertainty of each task, a method was proposed in\cite{Kendall2017MultitaskLU} that takes into account the importance of each task. This method, not only provides a way to tune the importance of different tasks but also simultaneously learns the uncertainty for each task, which as shown in \cite{Foliadis2021ReliableDL} is beneficial for the DL-based position using CSI fingerprints.

Another approach aiming at improving the generalizability of NN models is meta-learning. With meta-learning, a model is trained on multiple tasks or environments such that the minimization of the loss function in an unseen task is done more efficiently. Training is done by considering a meta-level objective such as the average positioning error across the multiple environments \cite{Gao22,Owfi23}. Meta-learning aims at having a trained model that generalizes better not only across the trained tasks but also facilitates learning an unseen task with a lower number of training samples, in contrast to MTL which only aims at learning better the trained tasks. 

Motivated by the two-step approach of conventional positioning methods, i.e., with parameter extraction from the CSI in a first step and a position determination in a second step, a two-part model trained with multi-task learning and a meta-level objective has been recently proposed in \cite{Foliadis2022MultiEnvironmentBM}. For UE positioning in different environments, i.e., different training tasks, different models are assumed with the first part of the models being common across all task and trained with CSI samples from all tasks (multi-task learning) aiming at minimizing the sum positioning error across all tasks (meta-level objective). The second part of the model of each task is trained to be environment specific by using only training data from each environment. The proposed approach in \cite{Foliadis2022MultiEnvironmentBM} is able to improve the positioning accuracy of the trained environments, as well as achieve a better generalizability when transferring the first part of the model and fine tuning the two-part model with CSI samples of a new environment.

\subsection{Contributions}

As proposed in \cite{Kendall2017MultitaskLU}, MTL benefits from uncertainty estimation. The training in MTL can be improved by determining the relative weighting of the losses of each task based on the associated uncertainty estimate \cite{Kendall2017MultitaskLU}. For this reason, in this paper we combine the results from \cite{Foliadis2021ReliableDL} and \cite{Foliadis2022MultiEnvironmentBM} to benefit from the MTL of different positioning tasks and from late fusion using uncertainty estimation. For a setup with multiple BSs and considering the positioning of a UE using each BS as a separate task, we show that employing a MTL scheme with uncertainty estimation and late fusion achieves high positioning accuracy. Additionally, even though this is outside the scope of the current paper, it was shown in \cite{Foliadis2022MultiEnvironmentBM} that a model trained with the MTL scheme can be further used for transfer learning in a new environment, reducing the time and amount of data that needs to be gathered.

Moreover, we extend the work in \cite{Foliadis2021ReliableDL} by employing a method described in \cite{Kumar2007AMF} for sensor fusion that takes into account the possibility that one or more sensors may be spurious. In the case of DL-based positioning, a model estimate could be spurious if the purported uncertainty is low but the real error is high. We employ this method in a late fusion scheme and show that it is beneficial in improving the positioning accuracy especially in dynamic environments.

Lastly, we aim not only to minimize the positioning error, but also evaluate the reliability of the uncertainty estimation. It would be beneficial if the estimated uncertainty truly reflects the model's uncertainty about the current measurement, such that a high uncertainty should indicate high positioning error and vice-versa. To evaluate the quality of uncertainty estimates we consider the area under sparsification error (AUSE) metric \cite{Ilg2018UncertaintyEA} \cite{Salihu2021TowardsSU}. In addition to the AUSE metric, we evaluate the integrity of the positioning results with respect to the integrity risk (IR)  which is used in global navigation satellite system (GNSS) applications and has been recently proposed in the Third Generation Partnership Project (3GPP) as a positioning key performance indicator (KPI) for 5G positioning \cite{3gpp}.

The paper is structured as follows. In Section II the considered system model is described along with the different types of fusion and the MTL scheme is introduced. In Section III the simulation setup is described and the DL-model structure. The results and conclusion are then presented in sections IV and V respectively.

\section{System Model}
\label{sec:system_model}

We consider an uplink setup with $N_B$ BSs each with $N_R$ receive antennas and a single transmit antenna at the UE. The UE transmits a reference signal on $N_C$ subcarriers within an orthogonal frequency division multiplexing (OFDM) symbol. The received uplink signal is used to estimate the CSI matrix between UE and each BS. The estimated channel at the $n$-th BS over the $N_C$ subcarriers is described as:
\begin{equation}
	\tilde{\boldsymbol{H}_n} = [ \tilde{\boldsymbol{h}}_0^n, \tilde{\boldsymbol{h}}_1^n,  \ldots, \tilde{\boldsymbol{h}}_{N_C-1}^n ] \in \mathbb{C}^{\BF{N_R \times N_C}},
\end{equation}
where $\tilde{\boldsymbol{h}}_l^n \in \mathbb{C}^{N_R \times 1}$ is a column vector that describes the estimated uplink channel between the UE and the $N_R$ antennas of the $n$-th BS at the $l$-th subcarrier. The estimated channels can be considered as a unique fingerprint of the position of the UE and depend on the multipath between the UE and each BS. To transform the raw complex CSI data to meaningful inputs for the NN, we stack the matrices $\Re{\tilde{\{\boldsymbol{H}_n\}}}$ and $\Im{\tilde{\{\boldsymbol{H}_n\}}}$ in the third dimension to obtain a new real-valued 3D matrix $\boldsymbol{H}_n \in \mathbb{R}^{N_R \times N_C \times 2}$. The symbols $\Re{\{\cdot\}}$ and $\Im{\{\cdot\}}$ denote the real and imaginary values of each of the matrix elements respectively. The values of each matrix are then normalized in the range $[0, 1]$. This transformation is a widely adopted practice in the literature for AI positioning using fingerprints as it allows the network to learn from both the magnitude and phase information, which are crucial for exploiting the multipath propagation effects captured by CSI\cite{Zhang2022CSIFingerprintingIL, Bast20, Foliadis21}. We input the matrix $\boldsymbol{H}_n$ to the DL-model without applying manual feature engineering and we leverage the model's ability to autonomously learn relevant features from the data \cite{Goodfellow-et-al-2016}.

\begin{figure}[b]
	\centering
	\hspace*{0cm}
	\includegraphics[scale=0.8]{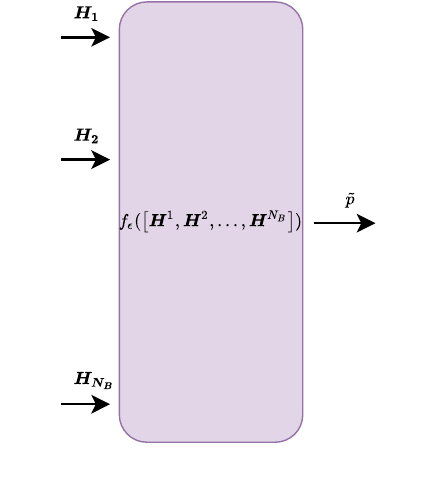}
	\caption{Early Fusion}
	\label{fig:early_fusion}
\end{figure}

\begin{figure}[b]
	\centering
	\hspace*{-0cm}
	\includegraphics[scale=0.7]{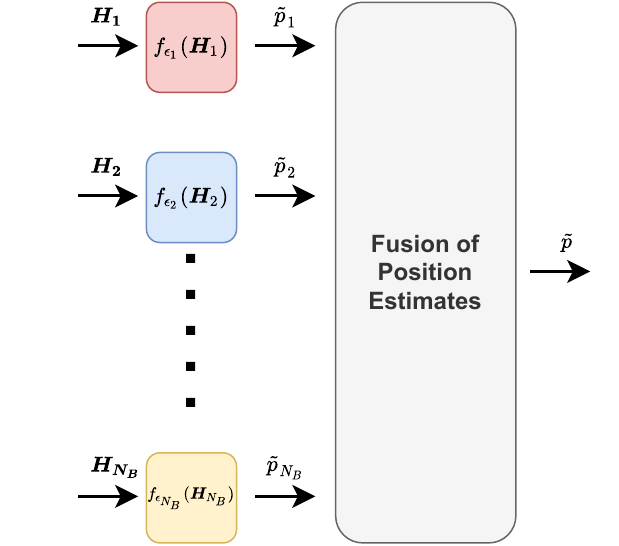}
	\caption{Late Fusion}
	\label{fig:late_fusion}
\end{figure}

\subsection{DL based positioning with fingerprints}

Deep learning based localization using CSI fingerprints as inputs consists of two phases, namely the training and the deployment phase, which are often alternatively termed as offline and online phases, respectively. During the training phase, CSI fingerprints are collected throughout the area of interest along with a label corresponding to the UE position associated with each CSI fingerprint. In order to collect fingerprints along with their labels, the use of positioning reference units (PRUs) can be employed, which consist of a device with known position, i.e., obtained with another positioning method or with sensors\cite{3gpp_tr_38_843}. Without  loss of generality, we assume that the UEs lie on a two dimensional plane. Subsequently, the CSI fingerprints and the position labels $\boldsymbol{p} =[x, y] \in \mathbb{R}^2$ are used to train the parameters $\epsilon$ of a neural network (NN) $f_\epsilon(.)$. Training is accomplished by minimizing the mean squared error between the position labels and the output of the NN with the labeled CSI fingerprints as input. Eventually, the trained NN is then used during the deployment phase to estimate the position $\tilde{{\boldsymbol{p}}}$ of a UE based on the estimated CSI fingerprint, where $\tilde{{\boldsymbol{p}}}  = [\tilde{x}, \tilde{y}]\in \mathbb{R}^2$ is the position estimate for the UE.

The key idea behind positioning with CSI fingerprints is that the CSI for each position is considered unique for that specific position. This stems from the fact that the channel between UE and BS is a rich source of information since it is influenced by various environmental factors such as walls objects or other obstacles. All this information is indirectly incorporated into the multipath propagation of the channel, which includes direct paths (LOS) and indirect paths (NLOS), and is extracted during the training phase of the NN. Consequently, positioning using fingerprints is part of modern positioning techniques such as \cite{Shahmansoori2017PositionAO}, which leverages both LOS and NLOS paths. Additionally, as shown in \cite{Foliadis2023DeepLB}, there is not necessarily a need for a LOS path at all since NLOS paths already contain information that can make the fingerprints unique and useful for positioning. The basic assumption is that the propagation environment should not significantly change between the training and deployment phases since that would degrade the performance of the NN.

Two different approaches for positioning using CSI fingerprints from multiple $N_B$ BSs can be considered \cite{Foliadis2021ReliableDL}, namely early and late fusion.

\subsubsection{Early Fusion}

In early fusion, a single DL-model is trained for the UE positioning, having as input the concatenation of the CSI fingerprints from all BSs, i.e., the single NN model  $f_\epsilon(\boldsymbol{H}) = \tilde{{\boldsymbol{p}}}$ where $\boldsymbol{H}=[\boldsymbol{H}_1, \boldsymbol{H}_2, ..., \boldsymbol{H}_{N_B}] \in \mathbb{R}^{(N_B \cdot N_R) \times N_C \times 2}$. Although this is a straightforward way to combine the information from all BSs and perform localization using a DL-model, it has some disadvantages. Firstly, a large signaling overhead is required in order to transmit the relevant CSI data to a central server which has the single NN model and second, if the setup changes (e.g. a BS is removed), then a new NN model has to be trained from scratch. A block diagram of early fusion is shown on Fig. \ref{fig:early_fusion}.

\subsubsection{Late Fusion}
With late fusion, a UE's position estimate is determined based on the CSI fingerprint at each BS. For this purpose, a separate NN model is trained at each BS based on the CSI at each BS as input. The parameters $\epsilon_n$ of the model $f_{\epsilon_n}(\boldsymbol{H}_n) = \tilde{{\boldsymbol{p}}}_n$ of the $n$-th BS are trained, with the CSI measurements $\boldsymbol{H}_n$ of that BS. During deployment, the position estimates obtained across all the BSs are appropriately combined to produce a single position estimate. This type of fusion is shown in Fig. \ref{fig:late_fusion}. We refer to this type of late fusion as, single task learning (STL) late fusion, since each DL-model is trained on a single specific task, namely the task of positioning based on CSI data from a specific BS.

Compared to early fusion, this method necessitates much less traffic for the network. The reason for that is that only the output of each of the models needs to be collected instead of the whole CSI fingerprints as in the case of early fusion. On the other hand, an appropriate model for the combination of the multiple estimates has to be developed. In this paper we built upon the work in \cite{Foliadis2021ReliableDL} and propose and compare methods to appropriately combine the position estimates considering the uncertainty.
\subsection{Uncertainty estimation}
Normally, for DL based positioning with CSI fingerprints, the parameters $\epsilon$ of the NN are optimized with respect to the mean squared error (MSE) loss:
\begin{equation}
	\mathcal{L}(f_\epsilon) = \mathcal{L}_x + \mathcal{L}_y,
	\label{eq:mse_loss}
\end{equation}
where $\mathcal{L}_x = E[|\tilde{x} - x|^2]$, $\mathcal{L}_y = E[|\tilde{y} - y|^2]$ and $E[.]$ denotes the mean value over the samples in the training set. 

The drawback of using such a loss function is that the model does not acquire any knowledge about the uncertainty that is present in the measurements. In the following, we discuss different types of uncertainties.

\subsubsection{Aleatoric Uncertainty}
\label{sec:aleatoric_UC}
The data dependent uncertainty is called aleatoric uncertainty and it reflects the uncertainty that a measurement has about the specific task. In a case of positioning using CSI fingerprints, a particular CSI measurement would have high uncertainty if it has low receive SNR for example. This type of aleatoric uncertainty, i.e. instance-dependent uncertainty, is called heteroscedastic uncertainty.
Since the aleatoric uncertainty in positioning using fingerprints is data-dependent, it can also be learned from the data. In \cite{Kendall2017WhatUD} a modification to the MSE loss was proposed in order to train a model to simultaneously calculate the position and the aleatoric uncertainty of the current position estimate. The loss function which shall be minimized with respect to the model parameters $\epsilon$ is the negative log-likelihood (NLL) function:
\begin{equation}
	\mathcal{L}'(f_\epsilon) = \frac{1}{2 {(\sigma_{x}^\alpha)}^2}\mathcal{L}_{x}+ \frac{1}{2 {(\sigma_{y}^\alpha)}^2}\mathcal{L}_{y} + \log({\sigma_{x}^\alpha}{\sigma_{y}^\alpha}),
	\label{eq:aleatoric_loss}
\end{equation}
where $\sigma_{x}^\alpha$ and $\sigma_{y}^\alpha$ are the aleatoric uncertainties for the outputs $x$ and $y$ respectively.

Subsequently, the output of the model has to be modified to include the learned aleatoric uncertainty $\boldsymbol{\sigma}^\alpha = [\sigma_{x}^\alpha$, $\sigma_{y}^\alpha]$ of $\tilde{\boldsymbol{p}} = [\tilde{x}, \tilde{y}]$, i.e., $f'_\epsilon(\boldsymbol{H}) = [\tilde{x}, \tilde{y}, \sigma_{x}^\alpha, \sigma_{y}^\alpha]$. The modified loss function $\mathcal{L}'(f_\epsilon)$ consists firstly of two regression terms that describe the inverse relationship between the MSE loss of each estimated coordinate and its corresponding aleatoric uncertainty. When the MSE loss for a particular coordinate cannot be minimized further, the model increases the respective aleatoric uncertainty to compensate. On the other hand the aleatoric uncertainty remains low for instances where the MSE loss is low. The last term is a regularization term that is used to limit the infinite increase of the outputs $\sigma_{x}^\alpha$ and $\sigma_{y}^\alpha$. 

\subsubsection{Epistemic Uncertainty}
Aleatoric uncertainty is not the only type of uncertainty present in a DL model. The other type is called epsitemic uncertainty and it accounts for uncertainty in the model's parameters \cite{Kendall2017WhatUD}. Estimates with high epistemic uncertainty indicate that the input comes from a distribution that was not learned by the model. In a DL localization model with fingerprints, the epistemic uncertainty would be high for a region in space where no data were collected or when a CSI measurement was corrupted. 

In \cite{Gal2016DropoutAA}, an approach for capturing a model's epistemic uncertainty called Monte Carlo dropout (MCD) was introduced. When employing dropout, random neurons in every weight layer of the deep learning model are deactivated with a  predefined probability. Typically, dropout is utilized solely for training purposes as a regularization technique \cite{Goodfellow-et-al-2016}, but with MCD, this same dropout probability is retained even during the deployment phase. Each successive forward pass through the deep DL model with MCD generates a unique configuration, and conducting multiple forward passes is akin to sampling from an approximate posterior distribution of the model's parameters given the dataset \cite{Gal2016DropoutAA}. The variance of the estimates from the different model configurations during these forward passes serves as an indicator of the model's epistemic uncertainty.


After $T$ forward passes, the combined aleatoric and epistemic uncertainty of the coordinate $x$ is \cite{Kendall2017WhatUD}:
\begin{equation}
	\sigma_x = \frac{1}{T} \sum_{t=1}^T\tilde{x}_t^2 - \Big(\frac{1}{T} \sum_{t=1}^T\tilde{x}_t\Big)^2 + \BF{\sigma_{x}^\alpha},
	\label{eq:MCD_variance}
\end{equation}
where $t$ indicates the current forward pass and $\tilde{x}_t$ indicates the estimate of the coordinate x at the $t$-th forward pass. The combined aleatoric and epistemic uncertainty for the $y$ position coordinate is calculated similarly.

Even though the epistemic uncertainty estimation is an efficient way for the DL model to report on its own knowledge about the current measurement, it is not always accurate. There are cases where the epistemic uncertainty is low but the mapping to the position is highly inaccurate. In those instances the model would provide a spurious estimate which has to be identified and eliminated from the fusion scheme. For the late fusion in \cite{Foliadis2021ReliableDL}, the method employed to combine the results from the different estimates is based on the assumption that each estimate follows a known Gaussian distribution, whose variance corresponds to the estimated combined aleatoric and epistemic uncertainty. As this assumptions does not always hold, we take into account such model inconsistencies by incorporating a method described in \cite{Kumar2007AMF} to fuse measurement from multiple sensors. The basic idea of this method is to weigh less the estimate that is most inconsistent with the other estimates. We should note that this method leverages multiple position and uncertainty estimates from different BSs, and therefore can only be employed in a late fusion scheme as described next.

In our setup we consider $N_B$ different models, corresponding to the models trained at each of the $N_B$ BSs. The authors of \cite{Kumar2007AMF} assume that the probability that a measurement from the $n$-th sensor $n \in [1, ..., N_B]$, is not spurious with probability 
\begin{equation}
	p_n = \exp({\frac{-(x - \tilde{x}_n)^2}{\alpha_n}})
\end{equation}
where $x$ is the true state, $\tilde{x}_n$ is the estimate of the $n$-th sensor and $\alpha_n$ is a parameter that depends on the variances of each separate model and the difference between the output of the $n$-th model with respect to other sensors:
\begin{equation}
	\alpha_n = \frac{b_n}{\prod_{ l=1, l \neq n}^{N_B}(\tilde{x}_n - \tilde{x}_l) ^2}
\end{equation} 
where $b_n$ is a hyperparameter. 

 From $\alpha_n$ we see that when the estimate of the $n$-th model is very different from the estimates of the other models, i.e., $\prod_{ l=1, l \neq n}^{N_B}(\tilde{x}_n - \tilde{x}_l) ^2 \to \infty$, then $\alpha_n \to 0$ and subsequently $p_n = 0$, meaning that the $n$-th estimate is definitely spurious. On the other hand when the $n$-th estimate largely agrees with the other estimates, i.e., $\prod_{ l=1, l \neq n}^{N_B}(\tilde{x}_n - \tilde{x}_l) ^2 \approx 0 $ then it follows that $p_n \approx 1$, which means that the $n$-th estimate is not spurious. To reflect this intuition he authors of \cite{Kumar2007AMF} show that the variance of the $n$-th model can be modified \cbend from $\sigma_{x, n}$ to:
\begin{equation}
	\sigma_{x, n}'^2 = \frac{\BF{\sigma_{x,n}}^2b_n^2}{b_n^2 - 2\sigma_{x, n}^2\prod_{ l=1, l \neq n}^{N_B}(\tilde{x}_n - \tilde{x}_l) ^2} 	
	\label{eq:SP_variance}
\end{equation}
where it must hold that $b_n^2 > 2\sigma_{x, n}^2\prod_{ l=1, l \neq n}^{N_B}(\tilde{x}_n - \tilde{x}_l) ^2$, so that $\sigma_n'> 0$. In our model we choose:
\begin{equation}
	b_n^2 = 2\sigma_{x, n}^2\prod_{ l=1, l \neq n}^{N_B}(\tilde{x}_n - \tilde{x}_l + \lambda) ^2
	\label{eq:beta_n}
\end{equation}
where $\lambda$ is a small valued hyperparameter.

By choosing the parameter $b_n$ as such, we make sure that when $\prod_{ l=1, l \neq n}^{N_B}(\tilde{x}_n - \tilde{x}_l) ^2 \to \infty$, then $\sigma_n' \to \infty$, reflecting very high uncertainty. On the other hand, when $\prod_{ l=1, l \neq n}^{N_B}(\tilde{x}_n - \tilde{x}_l) ^2 \approx 0 $, then $\sigma_n' \approx \sigma_n$. 
\subsection{Multi task learning}

\label{sec:MTL}
\begin{figure}[t]
	\centering
	\hspace*{-1.0cm}
	\includegraphics[scale=0.7]{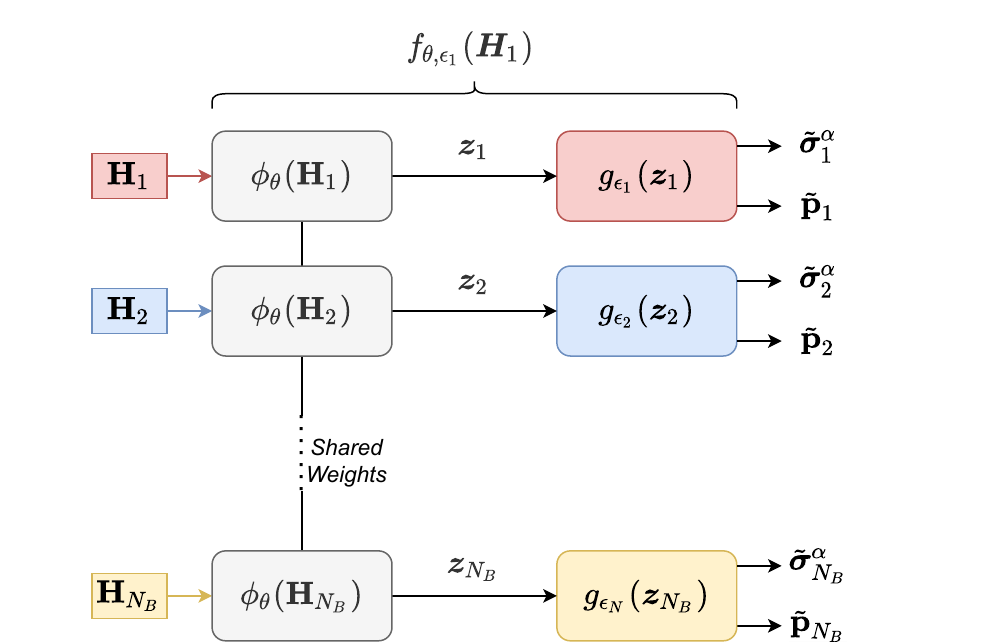}
	\caption{Two-part models with multi-task learning. Models' output comprise the UE's position estimate and the aleatoric uncertainy (when considered)}
	\label{fig:MTL}
\end{figure}

When considering the late fusion approach, i.e. a separate NN for each of the BSs, we propose sharing some parameters across models of different BSs as described in \cite{Foliadis2022MultiEnvironmentBM}. The $n$-th model which corresponds to the $n$-th BS $f_{\theta, \epsilon_n}(\boldsymbol{H}_n)$ is parametrized by the common parameters $\theta$ and the BS specific parameters $\epsilon_n$. By defining $f_{\theta, \epsilon_n}(\boldsymbol{H}_n) = g_{\epsilon_n}(\phi_\theta(\boldsymbol{H}_n))$ and training the parameters $\theta$ only on data from multiple BS we are forcing $\theta$ to be the same, regardless of the input data.  This means that we can deal with the training of the different models as a MTL scheme and jointly minimize the loss for the positioning with the model at each BS. MTL with parameter sharing allows for information flow between tasks which eventually may help each individual task \cite{Crawshaw2020MultiTaskLW}. The block diagram of a the MTL scheme considered in this work is shown in Fig. \ref{fig:MTL}.

This method of training is not possible when considering early fusion, since there is only one available model. Furthermore, when comparing STL late fusion to MTL late fusion we see that MTL requires the data from all BSs to be collected in order to train the models since the models share some parameters. For STL late fusion each BS trains its own model and then only shares the result of the model so there is no need to share input data between them. However, the fact that the models in MTL late fusion share parameters can enable training by means of federated learning \cite{McMahan2016CommunicationEfficientLO}. Federated learning refers to the technique whereby multiple nodes can train a model by partially training it locally and then sharing the model's parameters instead of sharing the data. This method can reduce data transfer requirements between nodes and also preserve privacy. Federated learning is also not applicable in the early fusion case since no single BS is able to do partial training on the model (see Fig. \ref{fig:early_fusion}) as the it needs CSI data from all BSs on its input to predict a single UE position.

The naive approach of optimizing a MTL scheme is to minimize a linear sum of the loss for each individual task, i.e. for the positioning at each BS. Thus, the loss for the training of the models in the MTL scheme would be would be:
\begin{equation}
	\mathcal{L}_{\text{MTL}}(f_{\theta, \epsilon_1}, f_{\theta, \epsilon_2}, ..., f_{\theta, \epsilon_{N_B}})=\sum_{n=1}^{N_B}\mathcal{L}_{n}(f_{\theta, \epsilon_n}),
	\label{eq:MTL_mse_loss}
\end{equation}
where $\mathcal{L}_{n}(f_{\theta, \epsilon_n})$ is the MSE loss of the $n$-th model as described in eq. \eqref{eq:mse_loss}. The authors of \cite{Kendall2017MultitaskLU} observed that by weighting each task appropriately the performance of each task can be greatly improved and proposed to set the weighting based on the aleatoric uncertainty \cite{Kendall2017WhatUD} of each task.

In the context of DL based localization using fingerprints each model is generates a 2-dimensional position estimate. By assuming that each output of the $n$-th DL model follows a Gaussian distribution $\mathcal{N}(\tilde{x}_n, {\sigma_{x, n}}^2 )$, and similarly for the $y$ variable, the authors of \cite{Kendall2017WhatUD} derive the minimization objective for the multi task learning as follows:
\begin{equation}
	\begin{split}
		\mathcal{L}'_{\text{MTL}}(f_{\theta, \epsilon_1}, f_{\theta, \epsilon_2}, ..., f_{\theta, \epsilon_{N_B}})=\sum_{n=1}^{N_B}\mathcal{L}'_{n}(f_{\theta, \epsilon_n}) \\ = \sum_{n=1}^{N_B}\bigg[\frac{1}{2 {(\sigma_{x, n}^\alpha)}^2}\mathcal{L}_{x, n}+ \frac{1}{2 {(\sigma_{y, n}^\alpha)}^2}\mathcal{L}_{y, n} + \log({\sigma_{x, n}^\alpha}{\sigma_{y, n}^\alpha})\bigg]
	\end{split}
	\label{eq:MTL_aleatoric_loss}
\end{equation}
where $\mathcal{L}'_n(f_{\theta, \epsilon_n})$ is the NLL loss of the $n$-th model as defined in \eqref{eq:aleatoric_loss} and it includes each model's aleatoric uncertainty $\sigma_{x, n}^\alpha$ and $\sigma_{y, n}^\alpha$ which is implicitly weighing the losses for each task.

\subsection{Late fusion with uncertainty estimation}

\label{sec:late_fusion}
The assumption that the outputs follow a Gaussian distribution, for which we have estimates of the mean value and variance, can be leveraged during the data fusion process. The $N_B$ estimates all refer to a single UE's position in a 2-dimensional plane and the fused probability distribution can be calculated using Bayesian inference\cite{Li2015ASO}. The resulting maximum likelihood (ML) fused estimate $\tilde{\boldsymbol{p}} = [\tilde{x}, \tilde{y}]$ $\sim$ $\mathcal{N}([\tilde{x}_\text{ML} , \tilde{y}_\text{ML} ],  \begin{bmatrix} \sigma_x^2, 0 \\ 0, \sigma_y^2  \end{bmatrix})$ has a variance of:
\begin{equation}
		\sigma_{x}^2= \frac{1}{\sum_{n=1}^{N_B}1/\sigma_{x, n}^{2}},\;\;\;  \sigma_{y}^2 =\frac{1}{ \sum_{n=1}^{N_B}1/\sigma_{y, n}^{2}}
\end{equation}
and a mean value:

\begin{equation}
	\begin{split}
	\tilde{x}_\text{ML} &= \frac{\sum_{n=1}^{N_B}{\tilde{x}_{n}}/{\sigma_{x, n}^{2}}}{1/\sigma_{x}^2},\\
	\tilde{y}_\text{ML} &= \frac{\sum_{n=1}^{N_B}{\tilde{y}_{n}}/{\sigma_{y, n}^{2}}}{1/\sigma_{y}^2}
	\label{eq:fusion}
	\end{split}
\end{equation}
and similarly for $\tilde{y}$. This type of estimate weighting is called inverse variance weighting. The value of $\sigma_{x, n}$ is given by \eqref{eq:MCD_variance} in the case of MCD uncertainty estimation, or \eqref{eq:SP_variance} when considering spurious estimates (SP). Additionally, if we don't take into account the uncertainty of each estimate we can consider $\sigma_{x, n} = \sigma_{y, n} = 1$ which this results to a simple averaging of the position estimates.

\subsection{Quality of uncertainty estimation}

After presenting different ways to calculate the uncertainty for each estimate, we now discuss how to assess the quality of these uncertainty estimates. As shown in the previous section, instead of providing a single position estimate, each model provides a different probability distribution for each individual input, for which the variance corresponds to the uncertainty. Normally to assess whether the output of the model indeed conforms to a probability distribution we would repeatedly produce samples from the model for a single input, calculate the empirical mean and variance and determine how close they are to the estimated model's distribution. However,
in the context of localization using CSI fingerprints we have at most a couple of CSI samples for a given location, therefore any empirical calculation would be unreliable. Instead we use a method to determine the reliability of the uncertainty estimation process which is called the area under sparsification error curve (AUSE).

\subsubsection{Area under sparsification error curve}

The idea behind this metric is to use the so-called sparsification plots as a quality metric and the sparsification error \cite{Ilg2018UncertaintyEA}. Before defining the sparsification error we first need to define the oracle error. We define an array of the errors of each position sample as:
\begin{equation}
	e = \left[||\tilde{x}_0 - x_0||_2, ||\tilde{x}_1 - x_1||_2, \ldots, ||\tilde{x}_{N_{\text{test}-1}} - x_{N_{\text{test}-1}} ||_2\right]
\end{equation}
where $\tilde{x}_i$ is the estimated value of the $i$-th sample and $x_i$ is the real value and $N_\text{test}$ is the number of samples in the test set. We also define the function $\text{sort}(.)$ which is used to sort the elements of an array in descending order, such that $e' = \text{sort}(e)$. With that, the oracle error can be defined as:
\begin{equation}
	\text{O}_N = \sqrt{\frac{\sum e'_{N:N_\text{test}-1}}{N_{\text{test}}-N}},
\end{equation}
where $e'_{N:N_{\text{test}}-1}=[e'_N, e'_{N+1}, \ldots, e'_{N_\text{test}-1}]$ and $1\leq N \leq N_\text{test}$.  The value $O_N$ is decreasing monotonically with $N$ since the errors are removed from the array $e'$ in decreasing order.

To define the sparsification error, we first define an uncertainty vector $s = [\sigma_0, \sigma_1, \ldots, \sigma_{N_\text{test}-1}]$ and also, the function $\text{argsort}_s(.)$ which sorts the elements of an array with respect to the descending order of the elements of the array $s$, and $e^S = \text{argsort}_s(e)$. With that, the sparsification error is defined as:
\begin{equation}
	S_N = \sqrt{\frac{\sum e^S_{N:N_\text{test}-1}}{N_{\text{test}}-N}},
\end{equation}
The sparsification error shows how much the estimated uncertainty coincides with the true errors on the test set. By removing gradually from the test set the estimates with the highest uncertainty and if the estimated uncertainty is of high quality the mean error $S_N $ should decrease monotonically when increasing $N$. We compare $S_N$ to $O_N$, by calculating the curve under the function $S_N-O_N$ for $N \in [1, N_{\text{test}}-1]$. The value AUSE is calculated as:
\begin{equation}
	\text{AUSE} = \frac{\sum_{N=1}^{N_{\text{test}-1}}S_N-O_N}{N_{\text{test}-1}}.
\end{equation}
A small value indicates that the sparsification error is close to the oracle error, meaning that the uncertainty estimation is a good indicator for the actual error in the test set.


\subsubsection{Integrity risk}

We additionally use the integrity risk (IR) metric which was recently proposed by 3GPP as a key performance indicator for positioning integrity \cite{3gpp}. Normally, if the uncertainty is high, the system should give a warning that the respective error is also high. The integrity risk is defined as the probability that the unknown positioning error exceeds an application specific alert limit (AL) without warning. The available information from each user is the position and the uncertainty estimate, therefore we define an indicator function $\mathbbm{1}_\text{AL}[||\boldsymbol{\sigma}_i||_2]$ which gives the aforementioned warning when the euclidean norm of the uncertainty $||\boldsymbol{\sigma}_i||_2$ vector of the $i$-th measurement is larger than some threshold $\gamma$ which corresponds to an position error equal to the AL:
\begin{equation}
	\mathbbm{1}_\text{AL}[||\boldsymbol{\sigma}_i||_2] = 
	\begin{cases}
		1  & \text{for  } ||\boldsymbol{\sigma}_i||_2 \leq \gamma \\
		0 & \text{for  } ||\boldsymbol{\sigma}_i||_2 > \gamma
	\end{cases}.
\end{equation}
The IR is then defined as:
\begin{equation}
	\text{IR} = \frac{\sum_{\{i | \mathbbm{1}_\text{AL}(||\boldsymbol{\sigma}_i||_2)\}}\mathbbm{1}_\text{AL}[e_i]}{N_\text{test}},
	\label{eq:IR}
\end{equation}
In words, the IR is calculated as the ratio of samples that exceeded the AL but no warning was given, to the total number of test samples. 

\subsection{Database description}
\label{sec:database_description}

\begin{figure}[t]
	\centering
	\hspace*{-0.8cm}
	\scalebox{.99}{\input{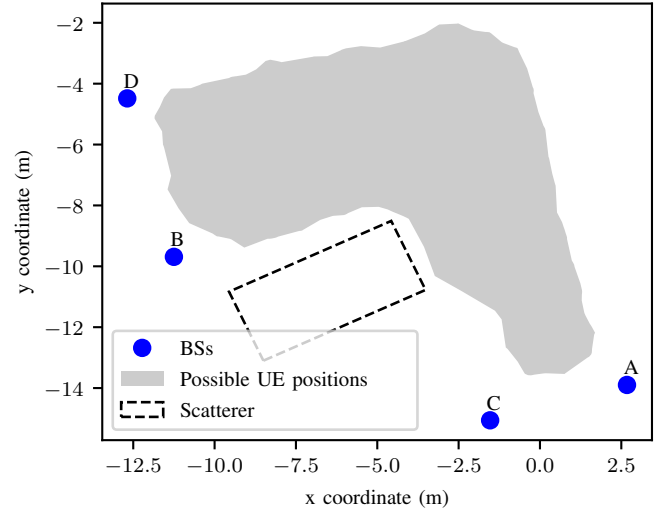}}
	\caption{Layout of considered industrial environment in Dichasus \cite{dichasus2021}}
	\label{fig:dichasus}
\end{figure}

To evaluate our proposals we use the Dichasus channel measurements described in \cite{dichasus2021},  that were collected at four antenna arrays distributed on the corners of an industrial area shown in Fig. \ref{fig:dichasus}. Each of the antenna arrays consists of a $4 \times 2$ uniform rectangular array (URA) with vertical and horizontal antenna spacing of half a wavelength. The measurements in \cite{dichasus2021} were collected with a single-antenna UE transmitting an OFDM signal in the uplink with a bandwidth of 50 MHz and with a pilot sent at every tenth subcarrier out of 1024 subcarriers, i.e., $Nc=103$. The carrier frequency is $f_C = 1.272$ GHz. The ground truth positions are measured with a tachymeter robotic total station, a very precise instrument that tracks the robot’s antenna with a laser with at least centimeter-level accuracy. This method aligns with the 3GPP work item \cite{3gpp-RP-234039}, where position labels for a PRU are available using a different positioning sensor.

In a real deployment of a positioning system using fingerprints, the CSI fingerprints are influenced by variations in the environment such as movement of objects or people throughout the area of interest. To model a change in the environment for a UE at a given position, i.e., between the training and deployment phase, we consider the attenuation of the strongest path from the UE  to a BS, i.e., due to blocking by a nearby person or object. Please note that the strongest path to a BS may correspond to a NLOS path, as some areas do not have a LOS to a BS.

\section{Simulation Setup}
\label{sec:simulation_setup}

\subsection{Dynamic scenario}
\label{sec:sim_dynamic}
We assume that the wireless signal propagates along a number of different paths to the BSs. Each path is associated with a complex gain, time-of-arrival and an angle-of-arrival. The totality of the paths and their parameters can fully describe the channel between a UE and a BS. In order to model the attenuation of the strongest path we first transform the channel from the antenna-subcarrier domain to the angle-delay domain by  means of the discrete Fourier transform (DFT) as described in \cite{Sun2018SingleSiteLB}.

After the matrix transformation to the angle-delay domain we identify the strongest path as the largest element of the matrix and we attenuate it by 20dB which corresponds to the attenuation effect caused by a human body at similar carrier frequency [43]. For a real system as the one considered, there is power leakage of each path to the neighboring elements although the matrix still remains mostly sparse. To take into account the power leakage we attenuate by the same amount a grid of $3 \times 3$ elements around the strongest path.  The matrix is then transformed back to the antenna-subcarrier domain, resulting in a modified CSI fingerprint due to the attenuation. Although a realistic blockage of the strongest path may not actually result in the same modified CSI fingerprint, our aim is to evaluate the performance of the considered approaches considered a change in the environment. This is regardless whether the change is realistic or not, as the point is that the CSI fingerprint learned for a given position has changed.
	
\begin{figure}[t]
	\centering
	\hspace{0pt}
	\includegraphics[scale=0.215]{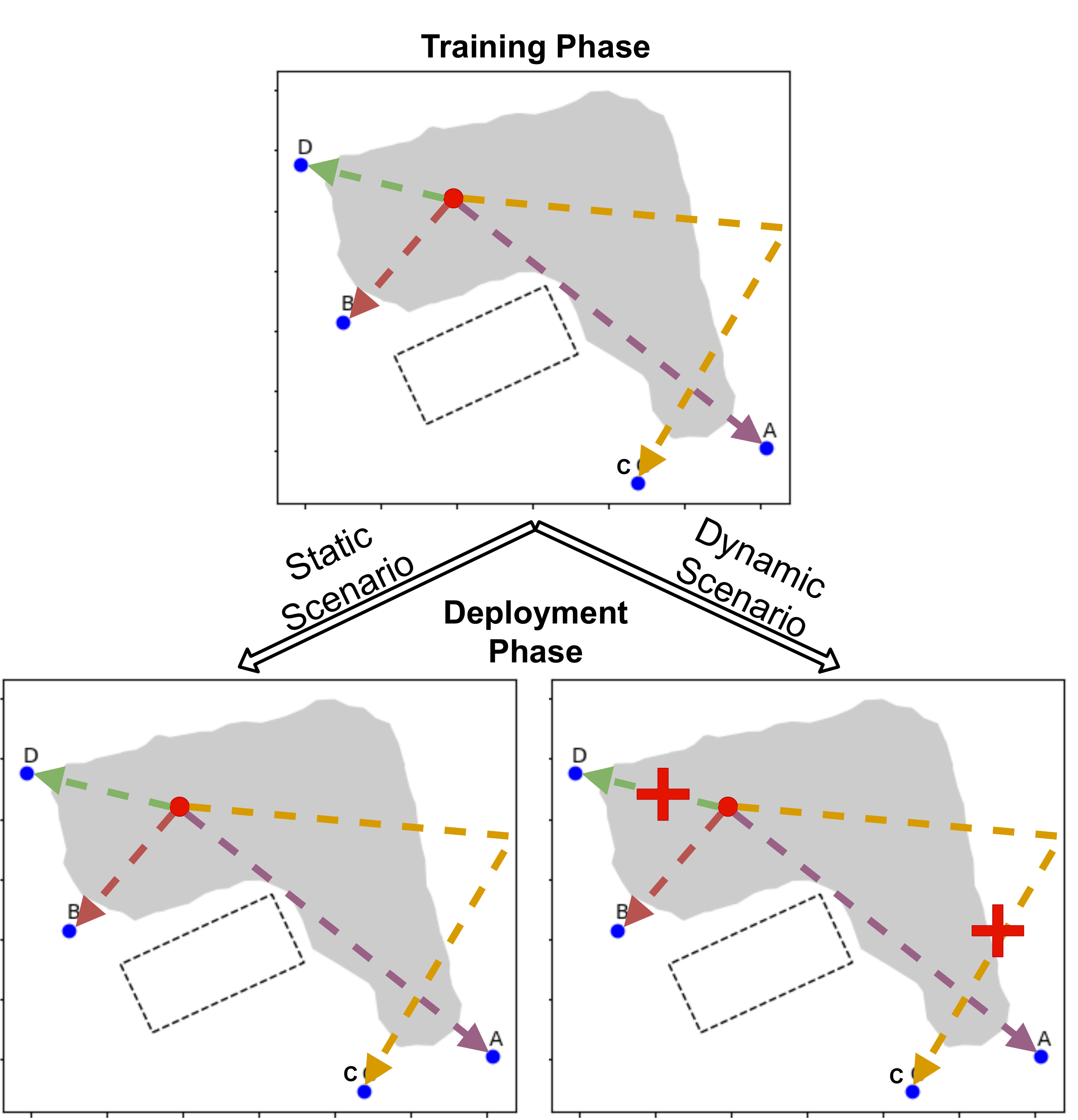}
	\caption{Static and dynamic scenarios example (strongest paths of 2 BSs attenuated). The UE is depicted as red dot inside the area of interest and the arrows indicate the strongest paths between UE and BSs.}
	\label{fig:Scenarios}
\end{figure}

Consequently, we consider scenarios with or without the above mentioned human body attenuation. When no attenuation is present we assume that there is no change in the environment between training and deployment phases, i.e. the environment is static as shown in Fig. \ref{fig:Scenarios}. Additionally we consider scenarios where attenuation affects the signal between UE and each of the BSs, but only in the deployment phase. This implies a change in the environment between the training and the deployment phases which we define as a dynamic scenario. We consider 4 different cases for the dynamic case considering a path attenuation from the UE to one, two, three or to all BSs. In the dynamic scenario example in Fig. \ref{fig:Scenarios} the strongest paths of BSs C and D are attenuated.
\begin{figure}[b]
	\centering
	\includegraphics[scale=0.6]{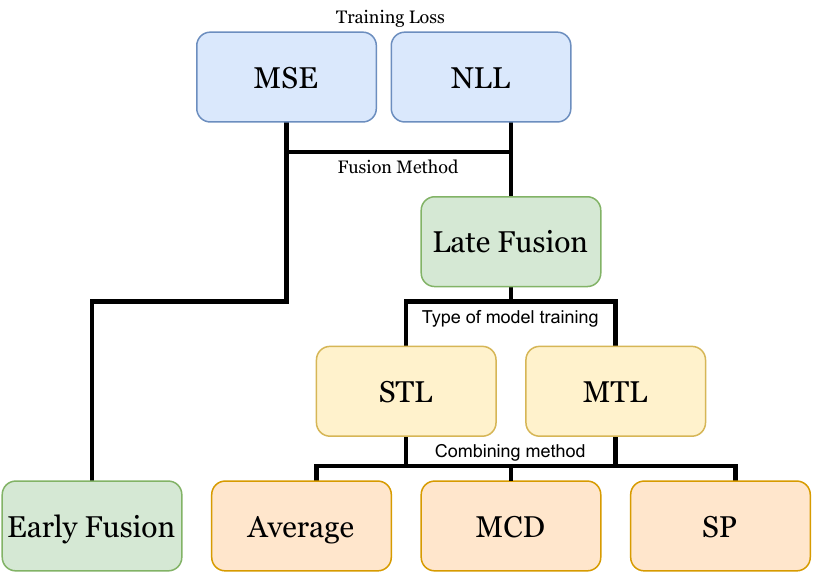}
	\caption{Overview of different fusion methods and training schemes}
	\label{fig:overview}
\end{figure}

Lastly, we compare all the different fusion approaches.
An overview of all the considered approaches is shown in Fig. \ref{fig:overview} considering different types of training loss, i.e., MSE or NLL loss, and different fusion methods, i.e., early or late fusion. For the late fusion specifically we have different types of model training.  Firstly, the STL late fusion which is the same method as described in \cite{Foliadis2021ReliableDL} where each BS corresponds to a single DL-model and each model is trained only on data from that BS. We also consider the MTL late fusion which assumes that the models of each BS share the parameters of their initial layers and are trained using the MTL scheme described in Section \ref{sec:MTL}. Finally we also consider different types of combining of the estimates from the multiple models, namely averaging, MCD or SP. With early fusion only one model is trained, with either MSE or NLL loss. 


\subsection{Neural Network Configuration}

\begin{figure}[t]
	\centering
	\hspace*{0cm}
	\includegraphics[scale=0.5]{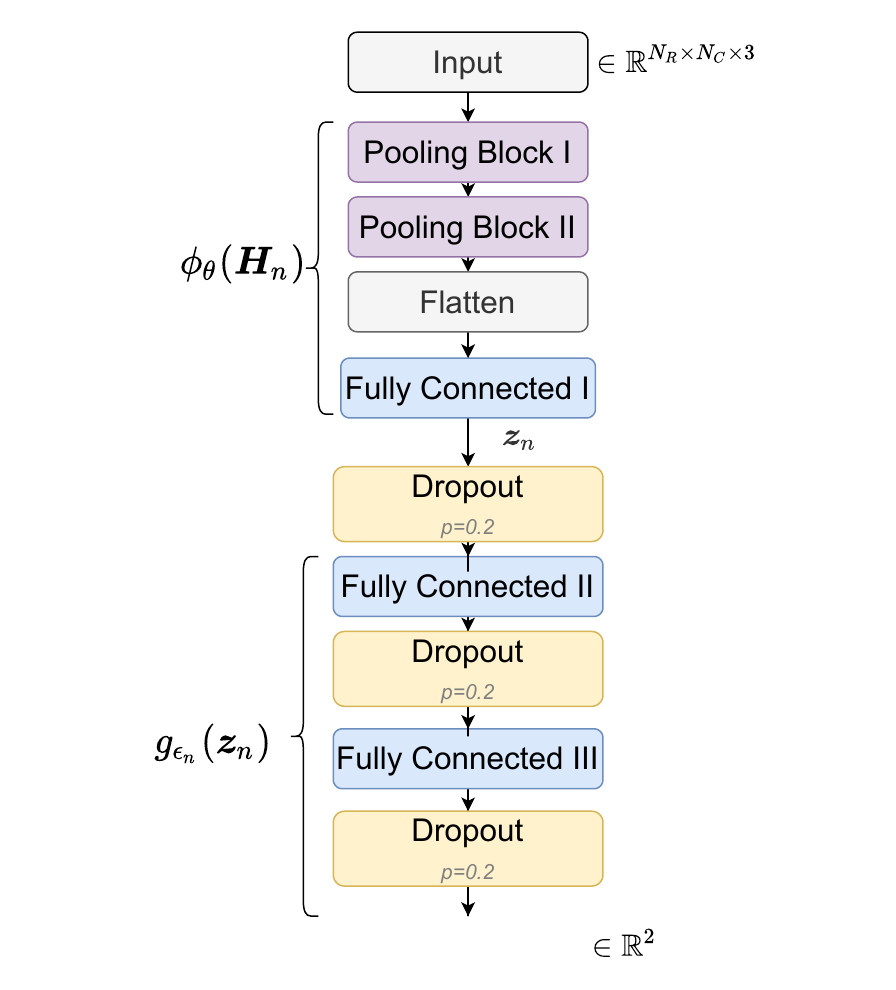}
	\caption{Complete DL model}
	\label{fig:nn_model}
\end{figure}

The considered neural network is shown in Fig. \ref{fig:nn_model}. In the MTL late fusion we consider that the models across BSs share the parameters of the first four blocks. Both early and late fusion use the same overall structure but with different input size, depending on the considered fingerprint.

\begin{figure}[b]
	\centering
	
	\includegraphics[scale=0.5]{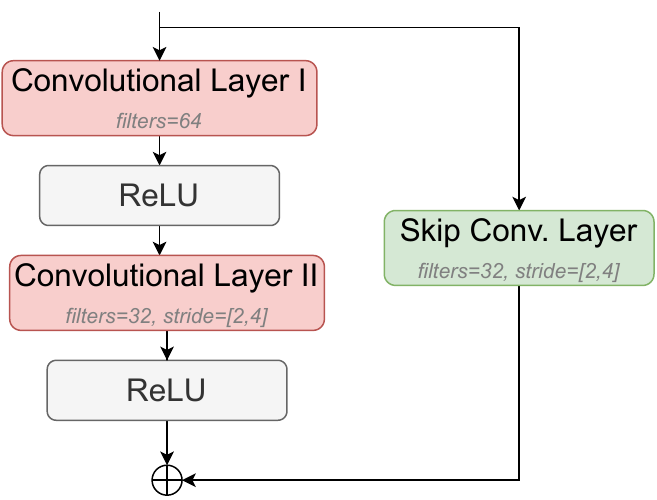}
	\caption{Pooling Block}
	\label{fig:pooling_block}
\end{figure}

The basis of the considered neural network is the convolutional layer as it has shown promising results for positioning using CSI fingerprints \cite{Foliadis2021ReliableDL, Foliadis2022MultiEnvironmentBM, Foliadis21, Chen17, Widmaier19, Wang17, Niitsoo18}. In general, the convolutional layer is followed by a pooling layer whose purpose is to downsample its input but recently \cite{Springenberg2014StrivingFS} it has been shown that using strided convolution instead of a pooling layer may improve the model's performance. Therefore, for the DL-model in this work we only use strided convolution and no pooling layers. A strided convolution can be thought of as a learned pooling layer, where the input is downsampled but the method of downsampling is learned during training \cite[chapter~9.5]{Goodfellow-et-al-2016}. 

Additionally, to further reduce information loss during downsampling, we implement the method of pooling blocks introduced in \cite{Lin2017RefineNet} which was also used for CSI based positioning in \cite{Zhang2022CSIFingerprintingIL}. In a pooling block, a convolutional layer doubles the number of learned convolutional filters before downsampling and then the spatial size is reduced by a strided convolution. A final convolutional layer is used to reduce again the number of learned convolutional filters to the original size. In this way, it is expected that the pooling block will learn to transfer the important information from the spatial dimension to the convolutional filters and preserve it.

\begin{table}[t!]
	\centering
	\caption{Fusion Schemes}
	\begin{tabular}{||c | c | c | c ||} 
		\hline
		\makecell{Training scheme}&\makecell{Input dim.} & \makecell{DL \\ models} & \makecell{Total \\ Parameters}  \\ [0.5ex] 
		\hline\hline
		Early fusion&$32 \times 103 \times 2$ & 1 & $357\ 954$ \\ 
		STL Late Fusion & $8 \times 103 \times 2$ & 4 & $743\ 688$ \\ 
		MTL Late Fusion & $8 \times 103 \times 2$ & 4 & $285\ 768$ \\ 
		\hline
	\end{tabular}
	\label{tab:fusion_schemes}
\end{table}
Lastly, in order to avoid any problems with vanishing gradients we employ skip connections \cite{He2016DeepResidual}. By combining all the aforementioned methods we create a pooling block which is shown in Fig. \ref{fig:pooling_block}. Two pooling blocks are placed one after the other and are followed by 3 dense layers with 128 neurons each and finally with a dense layer with 2 neurons that outputs the estimated 2-dimensional position. When considering the aleatoric uncertainty the last dense layer has 4 neurons which correspond to the 2-dimensional position plus the NLL loss for the $x$ and $y$ position coordinates. Each convolutional layer has 32 filters with a $3 \times 3$ kernel. The models are trained using the Adam optimizer with a batch size of 64 for 1000 epochs in total. The initial learning rate is $10^{-3}$ and is reduced to $10^{-4}$ after 100 epochs of no improvement of the validation loss.

Depending on the method used, i.e., early fusion, STL late fusion or MTL late fusion, the total number of model parameters are different. For the early fusion method, the input's dimension is $32 \times 103 \times 2$ which results from the CSI over all 32 antennas, i.e. 4 BSs with 8 antennas each, and 103 subcarriers. Therefore the total number of parameters in the early fusion case is $357\, 954$ parameters. When using STL late fusion , the CSI over the 8 antennas of a BS is used as input to each model. As the input dimensionality is then  $8 \times 103 \times 2$, the number of parameters per model is reduced to $185\, 922$. Since we have 4 models the total number of parameters is $743\, 688$. Lastly when jointly training the multiple models in a MTL scheme some parameters are shared so in this case the total number of parameters is $285\, 768$. The different configurations are summarized in Table \ref{tab:fusion_schemes}.

\section{Simulation Results}
We test the different proposed schemes using the Dichasus database \cite{dichasus2021} in the deployment area shown in Fig. \ref{fig:dichasus} by incrementally increasing the number of training samples $N_{\text{train}}$ from $N_{\text{train}}=10\, 000$ to $N_{\text{train}}=60\, 000$. The validation set is 10\% of the training set i.e. $N_{\text{val}} \in [1000, 6000]$. The hyperparameter $\lambda$ in Eq. (8) is chosen as 0.01. We compare the different schemes with respect to the mean error (ME), which is given by the mean euclidean distance between the estimated position and the true position in the test set. The number of test samples is $N_{\text{test}}=59\, 137$, regardless of the number of training and validation samples. Furthermore we compare the quality of the uncertainty estimation of the different schemes by the AUSE metric and the IR.

\subsection{Static scenario}
\begin{figure}
	\centering
	\input{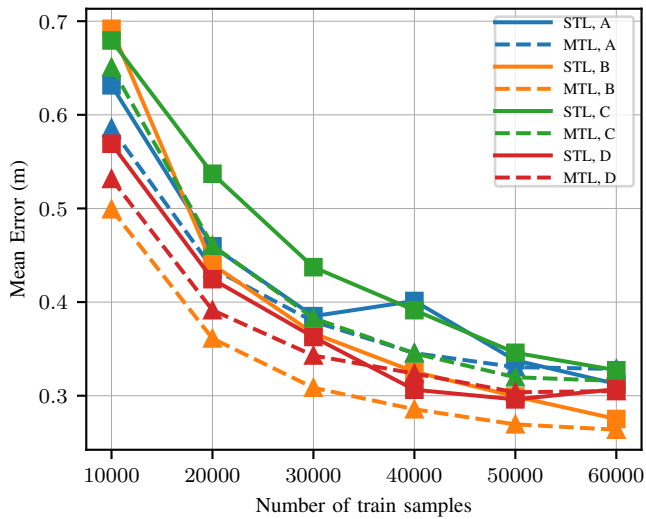}
	\caption{Comparison of ME of each BS in a static scenario when using STL or MTL scheme and MSE loss.}	
	\label{fig:error_static_bs_mse_S_mse_J}
\end{figure}

Initially we compare the different results for a static environment, i.e., when there is no change between training and deployment phases. Before comparing the different fusion approaches listed in Table \ref{tab:fusion_schemes}, we first show the gain of the training performed with the MTL training on the late fusion approach. For both the STL and MTL late fusion approaches, one model is trained at each BS by using the MSE loss in Eq. \eqref{eq:mse_loss} as the objective function. However, for the STL late fusion the model at each BS is trained only with data from that BS, while for MTL late fusion the first part of the model at each BS is trained with data from all 4 BSs. In the following, when mentioning MTL late fusion, we refer to joint training of the first common part of the models. Fig. \ref{fig:error_static_bs_mse_S_mse_J} shows the  performance of the models at each BS with the STL and MTL. The gain of joint training can clearly be seen across the models at each BS. By training the models jointly, their common part incorporates information from multiple BSs effectively increasing its training size. Thus, for a late fusion scheme, joint training the models at each BS (using MTL), instead of separately (STL), leads to a decreased ME at each BS. 

\begin{figure}
	\centering
	\input{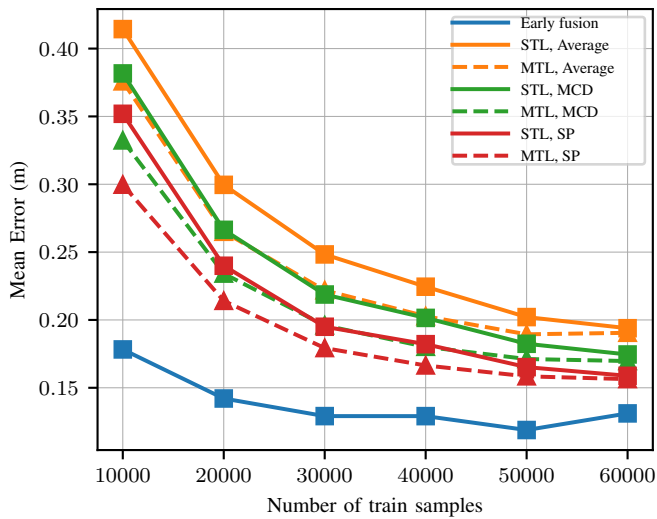}
	\caption{Comparison of ME for different fusion methods in a static scenario when training using STL or MTL scheme and MSE loss.}
	\label{fig:error_static_fusion_mse_S_mse_J}
\end{figure}
Next, we compare the different fusion schemes with respect to the ME in the test set. Specifically for both STL and MTL late fusion we consider three different methods of combining the estimates of the multiple models as described in section \ref{sec:late_fusion}. The first combining method is averaging where a simple average of all the model outputs is performed to calculate the overall estimate. The second combining method uses MCD-based combining where the variance of each estimate is estimated using the MCD method shown in eq. \eqref{eq:MCD_variance}. Lastly, we also consider the SP-based combining where the variance, shown in Eq \eqref{eq:SP_variance}, of the different estimates is modified by taking into account spurious measurements. The fused estimate for both MCD and SP is calculated using eq. \eqref{eq:fusion}. From Fig. \ref{fig:error_static_fusion_mse_S_mse_J} we can observe that using early fusion outperforms all late fusion methods in a static environment, i.e., when there are no changes in the environment between the training and the deployment phase. This is in contrast to the result from \cite{Foliadis2021ReliableDL}, where it was shown that late fusion outperformed early fusion in the static case. The difference in the conclusion of the results of \cite{Foliadis2021ReliableDL} and the ones shown in Fig. \ref{fig:error_static_fusion_mse_S_mse_J} can be explained due to the different considered environments. Whereas in \cite{Foliadis2021ReliableDL} there is always a LOS between the UE and each BS, in this work the link between the UE and each BS can either be LOS or NLOS. 

\begin{figure}
	\centering
	\input{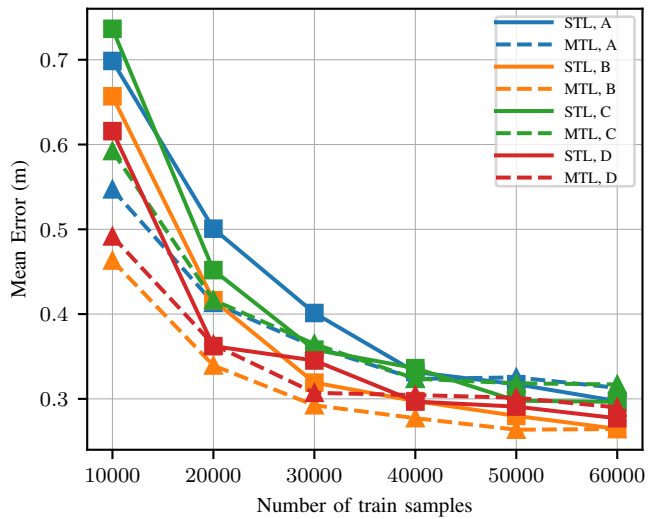}
	\caption{Comparison of ME of each BS in a static scenario when using STL or MTL scheme and NLL loss.}
	\label{fig:error_static_bs_al_S_al_J}
\end{figure}

We consider the performance of the different fusion approaches when the models are trained using the NLL loss function shown in Eq. \eqref{eq:aleatoric_loss}. Fig. \ref{fig:error_static_bs_al_S_al_J} shows the comparison of the models at each BS with separate and joint learning when using the NLL loss function. Similar to the MSE loss function case, there is an improvement in the performance for each model when training them jointly in a MTL scheme. As explained in \cite{Kendall2017MultitaskLU}, using a MTL scheme which enables joint late fusion, the aleatoric uncertainty is implicitly used as a learned weighting parameter between the losses corresponding to each task, i.e. the positioning at each BS, which can increase the performance for each task. 

\begin{figure}
	\centering
	\input{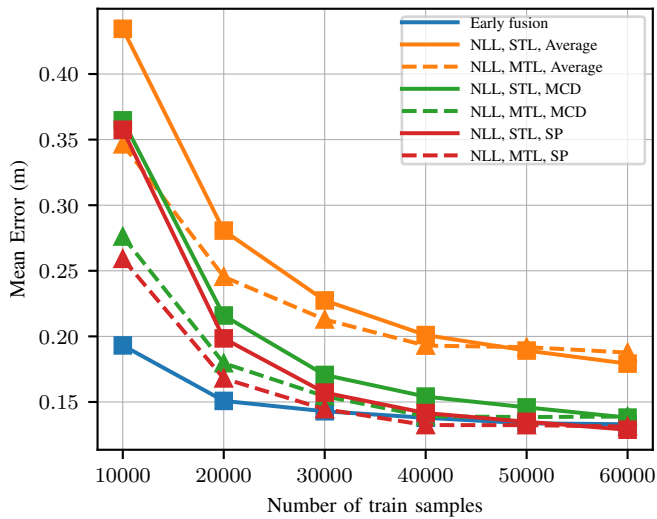}
	\caption{Comparison of ME for different fusion methods in a static scenario when using STL or MTL scheme and NLL loss.}
	\label{fig:error_static_fusion_al_S_al_J}
	
\end{figure}

\begin{figure}
	\centering
	\input{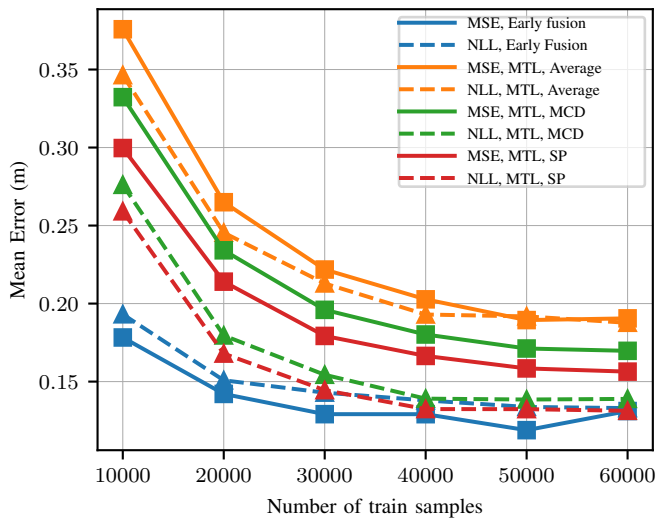}
	\caption{Comparison of ME for different fusion methods in a static scenario when training jointly with MSE or aleatoric loss}
	\label{fig:error_static_fusion_mse_J_al_J}
\end{figure}

We further compare the STL and MTL late fusion and the early fusion methods when training the models using the NLL loss. The results are shown in Fig. \ref{fig:error_static_fusion_al_S_al_J}. Similar to the results in Fig. \ref{fig:error_static_bs_al_S_al_J} when considering the MSE loss for the training, jointly training the models shows a significant improvement in performance compared to when training each model separately. This effect is particularly strong for a small number of training samples. We also see that even though the early fusion still outperforms the late fusion approaches for a small number of training samples, this is no longer the case when the number of training samples increases, i.e. the late fusion methods perform similar or better than the early fusion method.
Similar to when training with the MSE loss, the late fusion with averaging is the worst performing option.

Additionally, we compare the different fusion methods when training the models using the MSE \eqref{eq:mse_loss} or the NLL \eqref{eq:aleatoric_loss} loss functions. We see in Fig. \ref{fig:error_static_fusion_mse_J_al_J} that every late fusion scheme benefits when training the models based on the NLL function regardless of the number of training samples. Essentially, the inclusion of the aleatoric uncertainty improves the MTL late fusion schemes with MCD or SP-based combining, such that they are able to come close to the performance of the early fusion scheme, assuming an adequate number of training samples. For the early fusion scheme we see the opposite effect, namely that the inclusion of the aleatoric uncertainty during training impairs the performance, albeit slightly. The reason for this different behavior between late and early fusion may come from the fact the early fusion measurements always have some BSs having a LOS to the UE. This fact limits the usefulness of the aleatoric uncertainty since in a LOS measurement the uncertainty is anyway low. For the late fusion this is not the case since every BS may experience NLOS conditions which have high aleatoric uncertainty and positioning is then more challenging than the LOS case. As explained in section \ref{sec:aleatoric_UC}, the model prioritizes cases where the aleatoric uncertainty is low, i.e. LOS cases.

\setlength{\tabcolsep}{3.8pt}
\renewcommand{\arraystretch}{1.2}
\begin{table}[b!]
	\centering
	\caption{AUSE in static scenario}
	\begin{tabular}{||c | c | c | c | c | c||} 
		\hline
		\makecell{Training \\Loss}&\makecell{Early\\ Fusion} & \makecell{Separate \\MCD} & \makecell{Separate\\ SP} & \makecell{Joint \\MCD} & \makecell{Joint \\ SP} \\ [0.5ex] 
		\hline\hline
		MSE&0.51345 & 0.42626 & 0.38907 & 0.40984 & 0.38676\\ 
		NLL & 0.38434 & 0.33162 & 0.30825 & 0.32853 & 0.31335\\ 
		\hline
	\end{tabular}
	\label{tab:AUSE_static}
\end{table}

\begin{table}[b!]
	\centering
	\caption{AUSE in dynamic scenario}
	\hspace*{-0.11cm}
	\begin{tabular}{|| c | c | c | c | c | c | c||} 
		\hline
		\makecell{Blocked \\BSs}&\makecell{Training \\Loss}&\makecell{Early\\ Fusion} & \makecell{Separate \\MCD} & \makecell{Separate\\ SP} & \makecell{Joint \\MCD} & \makecell{Joint \\ SP} \\ [0.5ex] 
		\hline\hline
		\multirowcell{2}{A}&MSE&0.49605 & 0.42946 & 0.37902 & 0.41478 & 0.37673\\ 
		&NLL& 0.31853 & 0.33723 & 0.29964 & 0.31908 & 0.30472\\ 
		\hline
		\multirowcell{2}{A, B}&MSE&0.36338 & 0.42794 & 0.36777 & 0.40224 & 0.35451\\ 
		&NLL& 0.21424 & 0.32243 & 0.29882 & 0.28941 & 0.2768\\ 
		\hline
		\multirowcell{2}{A, B, C}&MSE&0.25143 & 0.53462 & 0.41211 & 0.41538 & 0.32425\\ 
		&NLL& 0.1995 & 0.36986 & 0.32064 & 0.31521 & 0.26985\\ 
		\hline
		\multirowcell{2}{A, B, C, D}&MSE&0.28687 & 0.52786 & 0.43868 & 0.38177 & 0.34691\\ 
		&NLL& 0.26857 & 0.44355 & 0.39339 & 0.38311 & 0.35556\\ 
		\hline
	\end{tabular}
	\label{tab:AUSE_dynamic}
\end{table}
Furthermore, we compare the quality of the uncertainty estimation in the different fusion methods in Table \ref{tab:AUSE_static}. 
During our evaluations, we noticed that the AUSE value remains mostly constant over training samples and therefore, we show the average over the training samples in Table \ref{tab:AUSE_static}. The averaging late fusion method it is not included in the table as it does not have uncertainty information. A lower AUSE value means that the sorting of the positioning errors across each measurement more closely corresponds with the sorting of the uncertainty, making the uncertainty a good indicator for the actual positioning error. We see from the table that for every fusion method, training using the NLL loss function improves the quality of the uncertainty estimates according to AUSE. This makes sense since there is no aleatoric uncertainty information when training MSE loss function, and instead only the epistemic uncertainty is used. 

\subsection{Dynamic Scenario}

\begin{figure}[b]
	\centering
	\hspace*{-0.5cm}
	\includegraphics[width=290pt, height=213pt]{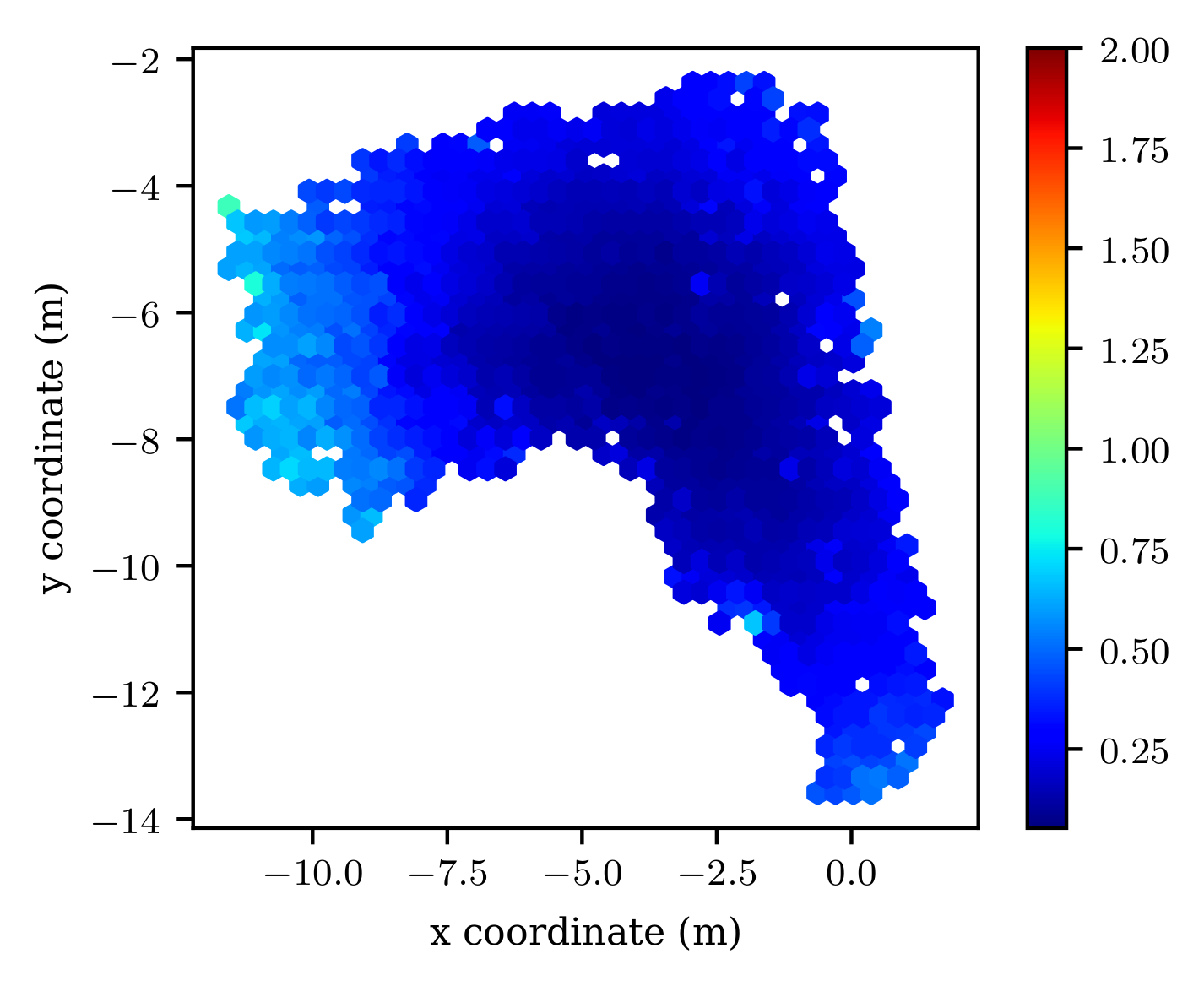}
	\caption{Uncertainty of model of BS A in a static scenario over all positions}
	\label{fig:sigma_A}
\end{figure}

\begin{figure}[b]
	\centering
	\hspace*{0cm}
	\includegraphics[width=240pt, height=213pt]{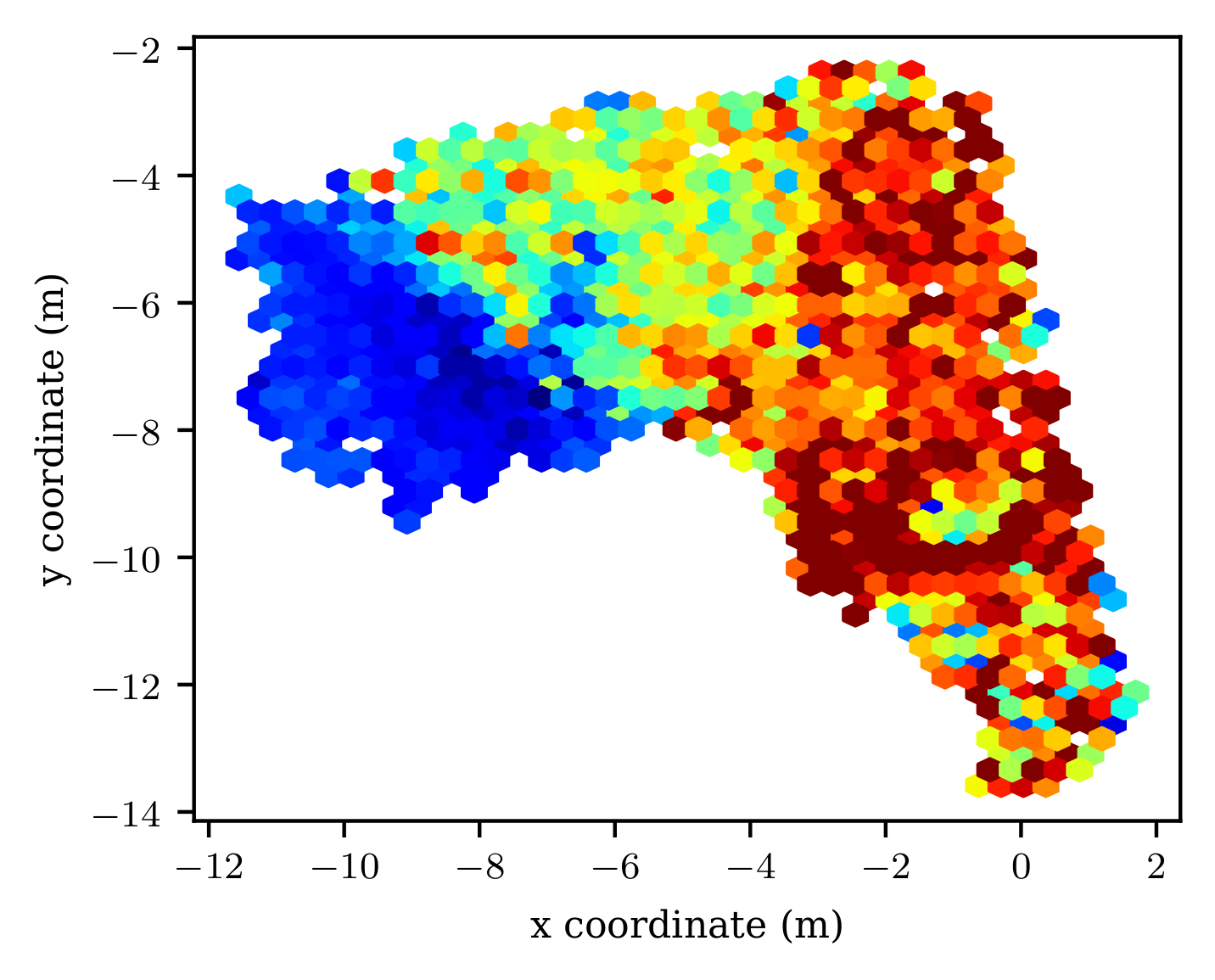}
	\caption{Uncertainty of model of BS A when it experiences a change}
	\label{fig:sigma_A_blockage}
\end{figure}

Next we explore the positioning in a dynamic scenario, where the channel between the UE and one or more BSs experiences a change between the training and deployment phase, i.e., a 20 dB attenuation of the strongest path as described in Sec. \ref{sec:sim_dynamic}. In the following, we refer to this attenuation of the strongest path to a given BS as a change. The effect of this change on the uncertainty of the estimates can be seen in Fig. \ref{fig:sigma_A_blockage}, which depicts the aleatoric uncertainty at BS A when the strongest path to BS is attenuated compared to the aleatoric uncertainty when no attenuation is considered shown in Fig. \ref{fig:sigma_A}. While the uncertainty in the static case remains more or less low throughout the area, i.e., around 0.25, when there is a change the uncertainty can be up to 8 times larger. We see that the most affected regions are the ones that there is a LOS path to the BS. This path includes most of the energy of the CSI thereby by reducing it the CSI is hugely affected. In the NLOS region we see that the uncertainty remains low since the position information is included in multiple paths, i.e., no single path contains most of the energy in the CSI fingerprint.

\begin{figure}
	\centering
	\input{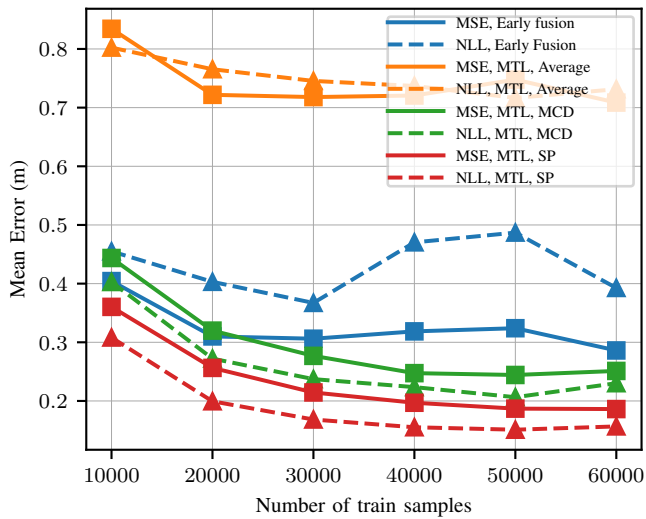}
	\caption{Comparison of ME for different fusion methods in a dynamic scenario when using MTL with MSE or NLL loss}
	\label{fig:error_dynamic_fusion_mse_J_al_J}
\end{figure}

Fig. \ref{fig:error_dynamic_fusion_mse_J_al_J} depicts the mean error of different fusion schemes when the channel between the UE and the BS A experiences the described change above in the deployment phase. Compared to the static case (see Fig. \ref{fig:error_static_fusion_mse_J_al_J}), we observe a huge degradation in performance for the late fusion with averaging, as well as a large degradation in performance for the early fusion case. As expected the late fusion schemes with MCD and SP combining perform much better than the other fusion methods. Specifically the SP-based late fusion method is able to outperform all others since it is able to most reliably disregard the spurious measurements due to the change in the channel to BS A.

\begin{figure}[t]
	\centering
	\input{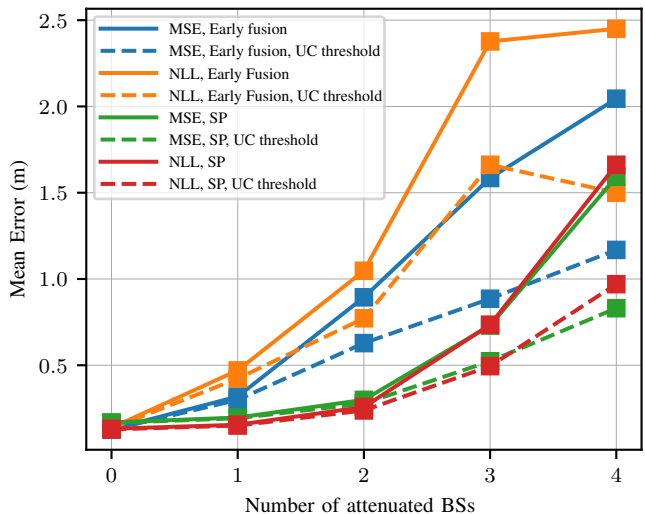}
	\caption{Comparison of the error of different error fusion for different number of blocked BSs, considering $40\,000$ training samples}
	\label{fig:error_blockage}
\end{figure}

Furthermore, we now consider a change in the channels between the UE and multiple BSs. In Fig. \ref{fig:error_blockage}, we depict the performance of early fusion and SP-based MTL late fusion when training with the MSE and NLL loss function as a function of the number of BSs with the attenuation of the strongest path, considering $40\, 000$ training samples. The solid lines show the error over all positions in the test set. As the uncertainty at different positions varies, we also propose to consider the performance for the positions when the uncertainty is below a threshold. By using logistic
regression in the static scenario we determine an uncertainty threshold for each method over which the positioning error is over 1m. Then for each method we exclude the measurements with uncertainty over this method-specific uncertainty threshold, and we see that the error decreases, depicted by the dashed lines in Fig. \ref{fig:error_blockage}. The difference is more pronounced for the larger number of BSs with a change and similarly in those cases more measurements are over the uncertainty threshold and therefore excluded. Interestingly, even though the difference between using the MSE loss or the NLL loss is relatively high for a small number of blocked BSs, with NLL loss training performing better, the difference becomes smaller when blocking more BSs. The reason for that is that when more BSs experience a change then the epistemic uncertainty dominates, since the measurements differ more from the static scenario.

Next, we investigate the reliability of the uncertainty estimates in a dynamic scenario with respect to the AUSE as shown in Table \ref{tab:AUSE_dynamic}. As in the static case, we provide the average over all training samples since similarly to the static case we noticed that the AUSE value remains mostly constant over the number of training samples. First we see that for almost every late fusion approach, training with NLL loss and MTL late fusion approach results in the most reliable uncertainty estimates. 

\begin{figure}[t]
	\centering
	\input{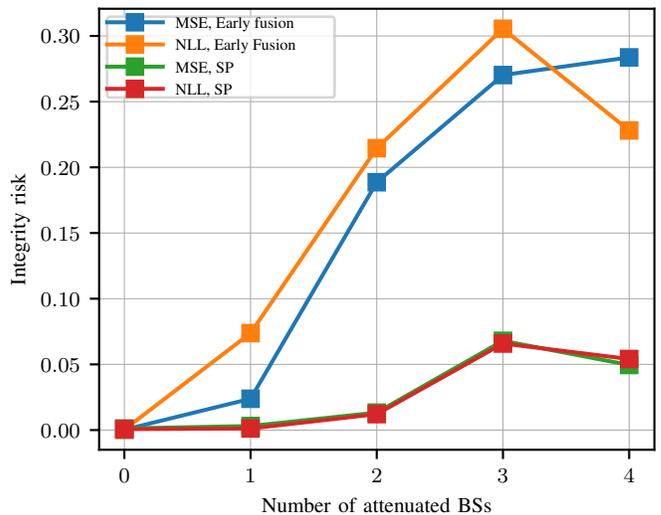}
	\caption{Integrity Risk}
	\label{fig:IR}
\end{figure}

Lastly, we depict in Fig. \ref{fig:IR} the integrity risk for $40\,000$ training samples considering a channel change to one or more BSs. The integrity risk is described in equation \eqref{eq:IR} and we consider $\text{AL} =1\text{m}$ and the threshold $\gamma$ is calculated using logistic regression in the static scenario for each fusion method. A low integrity risk shows that the uncertainty estimation can be used to identify estimates that exceed the alert limit. We see in Fig. \ref{fig:IR} that the SP method is the most reliable in this regard achieving an IR less than $7\%$ over all cases. On the other hand the early fusion methods exhibit a quite high integrity risk. The high integrity risk of those methods combined with the relative low AUSE shows that the early fusion methods exhibit overconfidence in their estimates. In other words, they are able to sort their errors based on their uncertainty, as indicated by the low AUSE, but the error corresponding to each uncertainty estimate is underestimated. This implies that with early fusion, it is assumed that some estimates are under the alert limit even though this is not the case.

We note that here we provide only a simple method to calculate the uncertainty threshold based on the uncertainty vector l2-norm and provide an IR value to show the effectiveness of the uncertainty estimation. Other methods to better calculate the threshold can be developed using both uncertainty vector elements and using other classification methods such as support vector machines (SVMs). Moreover, the considered metric does not indicate how many estimates are over the uncertainty threshold for each method which may be something that needs to be considered in some use cases.

	\section{Conclusions}


In this paper, we examined different fusion methods for positioning using deep learning and CSI fingerprints from multiple BSs. For early fusion, only one model is used for estimating the UE's position based on the CSI fingerprint across multiple base stations. For late fusion, one model per BS is employed and the overall UE's position is determined by combining the output of the models across the BSs. The performance of the trained models was evaluated considering a static scenario, where the channel between the training phase and the deployment phase remains the same, as well as in a dynamic scenario, where the channel between the UE and one or more BSs experience a change (attenuation of strongest path) in the deployment phase. While early fusion schemes may normally perform better in static scenarios, changes in the environment lead to a decrease positioning performance with early fusion, as the model is not able to adapt in a dynamic scenario. On the other hand, our results indicate that late fusion approaches are more robust to changes in the environment, which is an important aspect to be addressed for AI-based localization with CSI fingerprints in real deployments. Among the different considered late fusion approaches, we have shown the advantage of multi-task learning, by jointly training shared parameters of the models across the base stations, where the common part of the models benefits from a larger number of training samples.

For the late fusion approaches, different methods for combining the positioning estimates from the BS models have also been investigated. In particular, we have considered simple averaging as well as combining based on considering uncertainty estimation, namely MCD and SP, where the output of the different models are weighted based on the learned aleatoric uncertainty. We show that fusing the multiple estimates based on their uncertainty not only improves the positioning accuracy in both a static and dynamic scenario but also ultimately gives more reliable uncertainty estimates. The reliability of the uncertainty estimates is determined in terms of AUSE, which considers whether the uncertainty corresponds to the real positioning error, and in terms of the IR,
which demonstrates a model's ability to discard unreliable estimates. Additionally we consider that some of the estimates
may be spurious, i.e., falsely indicate low uncertainty but with an actual large positioning error, and we employ a technique to
identify and disregard such estimates.

Overall, we show that late fusion scheme with multi task learning and uncertainty estimation is the most accurate and reliable in the considered scenarios. This holds also for the dynamic scenario, which is one of the main challenges limiting the deployment of AI-based localization with CSI fingerprints.

\bibliographystyle{IEEEtran}
\bibliography{IEEEabrv,References}
\end{document}